\documentclass[10pt,letterpaper]{article}
\usepackage[top=0.85in,footskip=0.75in,marginparwidth=2in]{geometry}

\usepackage[utf8]{inputenc}

\usepackage[english]{babel}
\usepackage{blindtext}
\usepackage{times}  
\usepackage{helvet}  
\usepackage{courier}  
\usepackage{graphicx} 
\usepackage{comment}
\usepackage[ruled,vlined]{algorithm2e}
\usepackage{amsmath}
\usepackage{booktabs}
\usepackage{multirow, multicol}
\usepackage{array}
\usepackage{times}
\usepackage{adjustbox}
\usepackage{array}
\usepackage{cellspace}
\usepackage{tabularx}
\usepackage{verbatim}
\usepackage{cite}

\newcolumntype{L}[1]{>{\raggedright\arraybackslash}p{#1}}
\newcolumntype{C}[1]{>{\centering\arraybackslash}p{#1}}
\newcolumntype{R}[1]{>{\raggedleft\arraybackslash}p{#1}}
 
\usepackage{subfig}

\begin{document}

\vspace*{0.35in}

\begin{flushleft}
{\Large
\textbf\newline{DNA data storage, sequencing data-carrying DNA}
}
\newline
\\
Jasmine Quah,
Omer Sella,
Thomas Heinis
\\
\bigskip
Imperial College London
\\
\bigskip
* o.sella@imperial.ac.uk

\end{flushleft}

\begin{abstract}
DNA is a leading candidate as the next archival storage media due to its density, durability and sustainability.
To read (and write) data DNA storage exploits technology that has been developed over decades to sequence naturally occurring DNA in the life sciences. To achieve higher accuracy for previously unseen, biological DNA, sequencing relies on extending and training deep machine learning models known as basecallers. This growth in model complexity requires substantial resources, both computational and data sets. It also eliminates the possibility of a compact read head for DNA as a storage medium.

We argue that we need to depart from blindly using sequencing models from the life sciences for DNA data storage. The difference is striking: for life science applications we have no control over the DNA, however, in the case of DNA data storage, we control how it is written, as well as the particular write head. More specifically, data-carrying DNA can be modulated and embedded with alignment markers and error correcting codes to guarantee higher fidelity and to carry out some of the work that the machine learning models perform.






In this paper, we study accuracy trade-offs between deep model size and error correcting codes. We show that, starting with a model size of 107MB, the reduced accuracy from model compression can be compensated by using simple error correcting codes in the DNA sequences. In our experiments, we show that a substantial reduction in the size of the model does not incur an undue penalty for the error correcting codes used, therefore paving the way for portable data-carrying DNA read head. Crucially, we show that through the joint use of model compression and error correcting codes, we achieve a higher read accuracy than without compression and error correction codes.


\end{abstract}

\section{Introduction}

\begin{figure}[h!]
\centering
\includegraphics[width=.45\columnwidth]{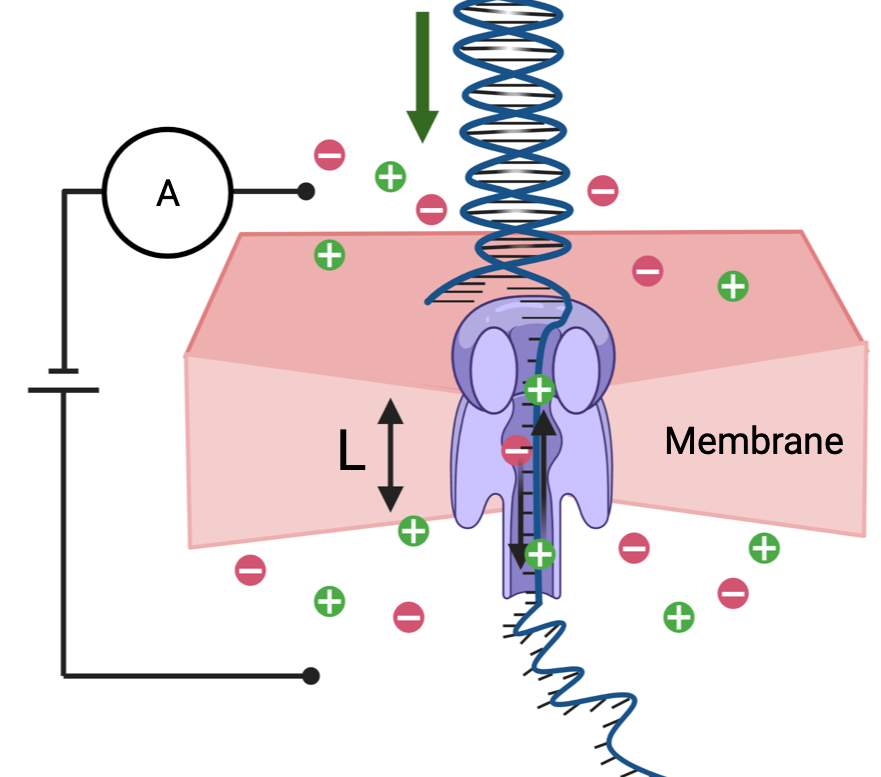}
\caption{A DNA molecule passes through a pore in a membrane, resulting in the modulation of an electric signal.}
\label{fig:nanpore}
\end{figure}

\begin{figure}[h!]
\centering
\includegraphics[width=.6\columnwidth]{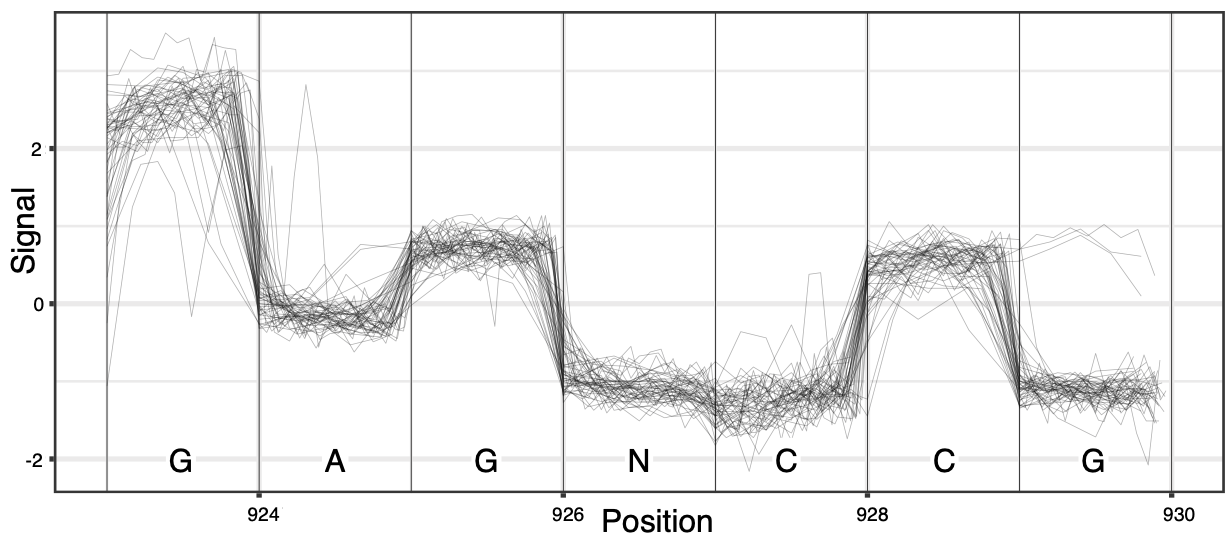}
\caption{Raw signal from Nanopore sequencing, showing the signal versus the nucleotides passing through the pore.}
\label{fig:squiggle}
\end{figure}

\begin{figure}[h!]
\centering
\includegraphics[width=0.8\columnwidth]{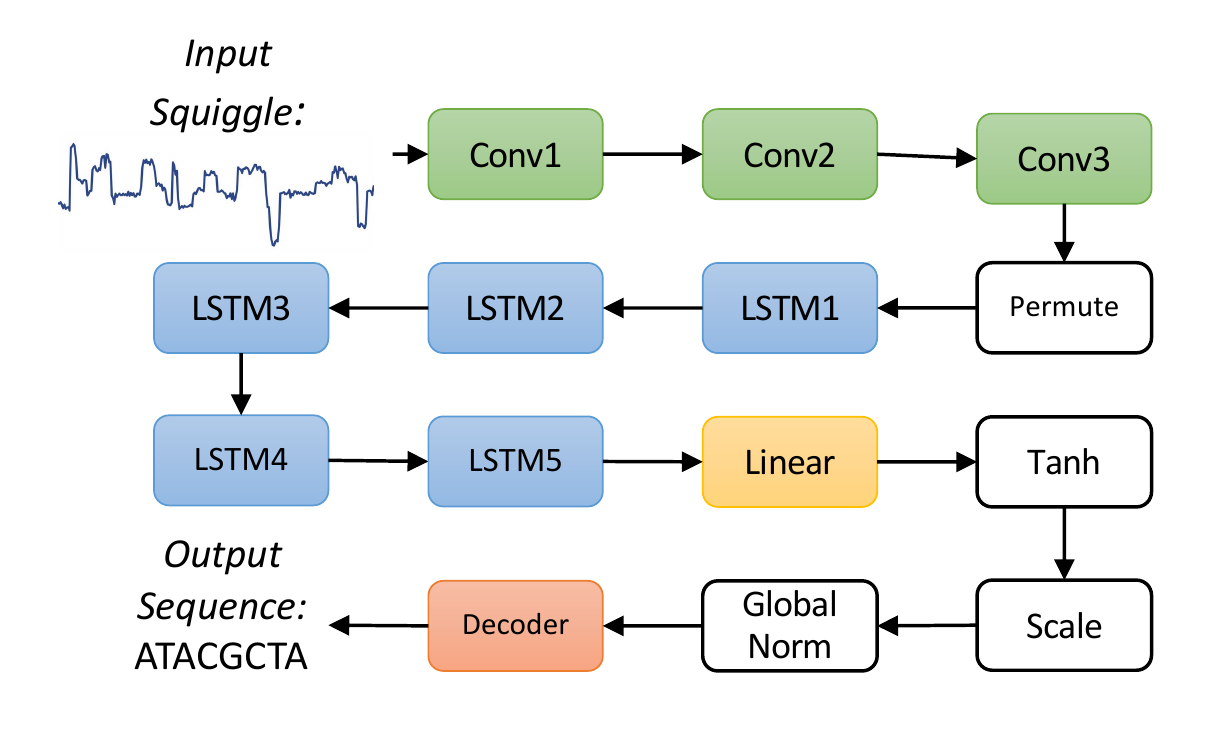}
\caption{The electrical signal is processed by a deep mode called a basecaller. The basecaller maps the signal to a sequence of bases (A,C,T,G). In this paper we use the Bonito architecture with model size of 107 MB and an accuracy of 97.317\%.}
\label{fig:bonitoArch}
\end{figure}

\begin{figure*}[ht]
  \begin{center}
    \includegraphics[width=.95\linewidth]{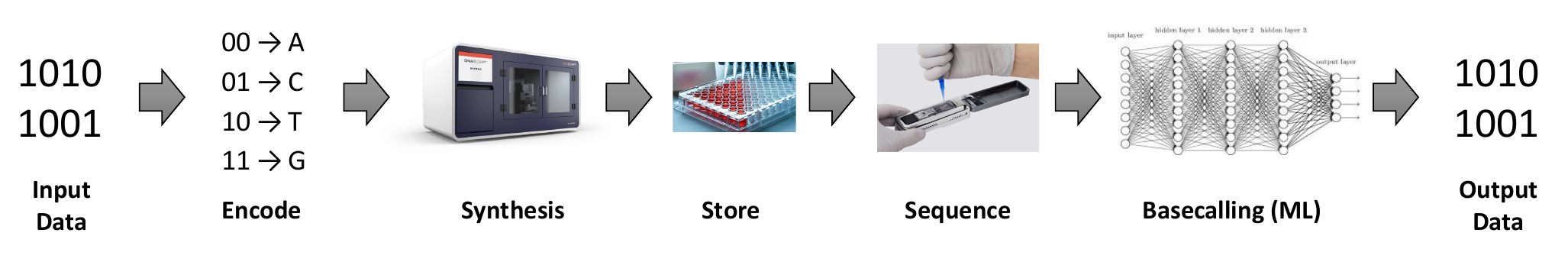}
    \caption{Binary input data is encoded, synthesised into DNA molecules which are stored. To read back, the DNA molecules are sequenced and decoded to obtain the initial data.}
    \label{fig:end2end}
  \end{center}
\end{figure*}

DNA data storage, where DNA is used as a medium to store arbitrary binary data, has been proposed as a sustainable solution to store humanity's data for centuries to come \cite{Goldman2013TowardsDNA}. DNA is extremely durable with a half-life of approximately 520 years~\cite{Allentoft2012TheFossils}. This means that data stored on DNA does not have to be migrated every 10-20 years due to the short-lived nature of traditional archival storage media, resulting in lower cost and less electronic waste. Furthermore, DNA is an incredibly dense storage medium with a theoretical capacity of 455 EB/g, i.e., 455 thousand Terabytes per gram~\cite{Church2012Next-generationDNA}, meaning 
that data centers built around DNA storage have the potential to be designed smaller, again resulting in lower cost of maintenance, land mass and energy. \\
However, several challenges regarding the read and write head as well as long-term storage have to be addressed before DNA data storage can become a viable alternative to traditional archival storage media and unleash its full potential.
On the read end, the focus of this paper, DNA data storage relies on sequencing, i.e., the ability to map a DNA molecule to a textual sequence of four symbols, A,C,T,G, corresponding to the DNA building blocks. This textual sequence is then converted back to a binary representation of the data. Modern sequencing has had many recent success stories \cite{nurk2021complete, miga2020telomere} as various architectures of deep neural networks have driven the accuracy of basecalling, i.e., the process of mapping signal to bases, higher \cite{wick2019performance}.\\
Retrieving the data back from DNA molecules to binary has so far leveraged traditional sequencing used for life sciences. Clearly, both data storage and Bioinformatics benefit from higher sequencing accuracy. So far major leaps in basecalling and sequencing were driven by the need of sequencing of living organisms, viruses~\cite{bull2020analytical} etc., using deeper and more complex machine learning models. For sequencing of data-carrying-DNA, however, this poses a problem. As basecallers grow deeper and contain more parameters they need to be trained on larger data sets. Furthermore, this growth in model complexity requires substantial resources and eliminates the possibility of a compact read head. In fact, while the hardware required to generate the input signal to the model is becoming smaller (Figure \ref{fig:end2end}) and more portable; it is the increasing size of the model that prevents the option of a portable DNA data reader.
Reducing the complexity and depth of basecalling models, and thus inference overhead, is therefore key. Model compression \cite{bucilua2006model} offers the chance to reduce hardware requirements but also reduces accuracy of the inference. When sequencing DNA from natural biological origin, this is not a viable technique, as we need as high accuracy as possible. Unlike use cases from the life sciences, however, data-carrying-DNA can be modulated and embedded with alignment markers and error correcting codes to guarantee higher fidelity and carry some of the work that the basecaller performs. In this paper we explore the trade-off between model size, its accuracy and error correction codes used. In our experiments we show that the model can be drastically reduced while still being able to retrieve the source data with high fidelity through the addition of minimal error correction codes in the sequence. Our contribution is a demonstration of the opportunities for co-design between error correcting codes and machine learning for data-carrying-DNA. This is a significant departure from the state-of-the-art which is still virtually exclusively driven by biological applications which cannot afford this luxury of co-design.


\section{Background}

\subsection{DNA Data Storage}
To store arbitrary binary data, a mapping is used to turn a binary sequence to a sequence whose content is made of four symbols, i.e.,  A,C,T,G. This is because a DNA molecule is made up of four distinct bases: Adenine, Thymine, Guanine, and Cytosine, designated A, T, G, C respectively. Through chemical or enzymatic synthesis, the resulting mapped sequence is physically created as a DNA molecule. The synthesised DNA molecules can then be stored for long periods of time.\\
DNA sequencing is used to read the information back. Two types of sequencing technologies have emerged: sequencing by synthesis (SBS) and Nanopore sequencing. Most SBS sequencers have a read-length limitation in the order of hundreds of nucleotides and must therefore go through significant molecular biological preparation including fragmentation (to agree with the read length) among others. The need for significant preparatory steps and limited read length for SBS sequencing is the reason why novel Nanopore sequencers are gaining traction, as they can sequence long DNA molecules without the need for fragmentation. In either case, DNA molecules are sequenced to obtain a raw signal (electrical, optical, etc.) which then has to be interpreted as a sequence of bases in a process called basecalling. In the case of nanopore sequencing, the signal, shown in Figure~\ref{fig:squiggle}, is the ion flow through the nanopore which is disturbed as individual bases of a DNA molecule pass through. The process is illustrated in Figure~\ref{fig:nanpore} where a single stranded DNA sequence passes through a nanopore while a signal is emitted. The resulting signal is commonly referred to as a squiggle.    
Basecalling is then used to call out individual bases which produces a DNA sequence, i.e.,: a sequence made from the symbols A,C,T,G. These are then inversely mapped back to a binary sequence. Comparing the source DNA sequence to the basecaller output, one may identify deletions, insertions or mismatches. For a survey of DNA data storage the reader is referred to \cite{heinis2019survey}.

\subsection{Basecalling Models}
Multiple basecallers have been developed to map raw nanopore signals to sequences~\cite{rang2018squiggle}. The Bonito basecaller is Oxford Nanopore Technologies' most recent basecaller, with the highest single read accuracy to date~\cite{bonito}. For this reason and because it is currently the most broadly used basecaller, we focus on it as a candidate for model compression. The original Bonito model, which consists of several blocks of convolutional layers with 6,649,613 network parameters, was heavily based on the Quartznet model~\cite{quartznet}. The most recent network architecture of Bonito is depicted in Figure \ref{fig:bonitoArch} and is made up of three convolutional layers, five LSTM layers, and one linear layer, with a total of 27,795,560 parameters. It has significantly more parameters than the original model, resulting in a size of 107MB.

\subsection{Error Correcting Codes}
\begin{figure}[h]
  \begin{center}
    \includegraphics[width=.65\columnwidth]{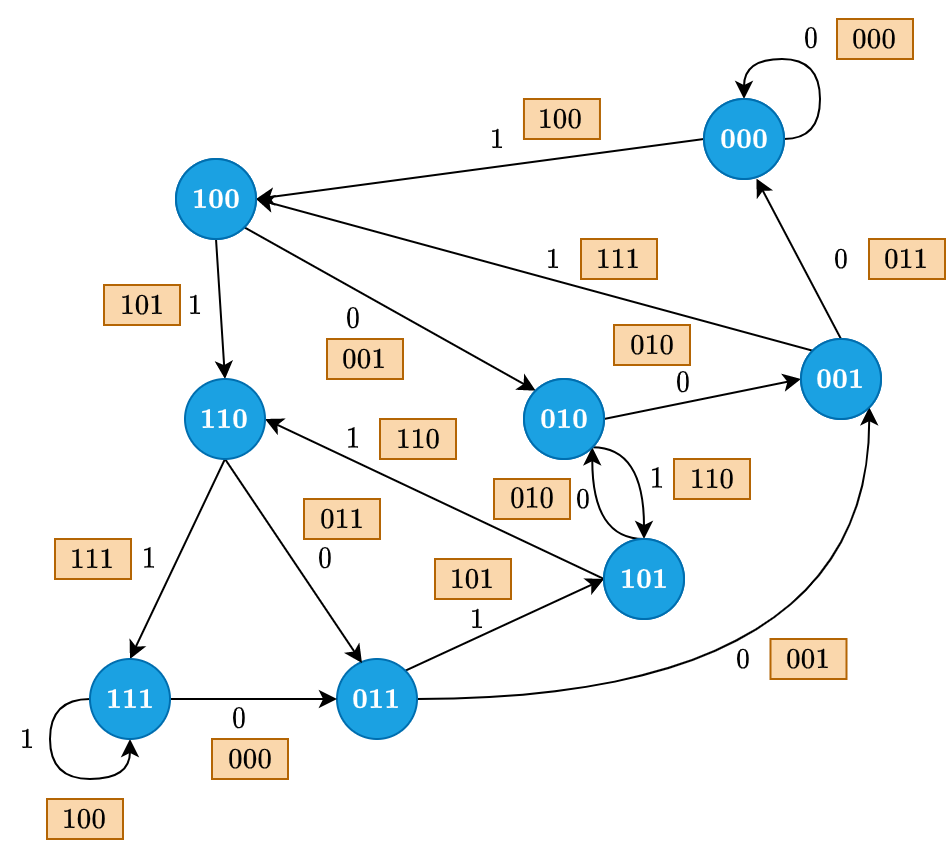}
    \caption{Finite state machine for 1/3 code used by encoder. The circled states represent the contents of the shift register, whilst the boxed bits are the output bits which form the encoded stream transmitted through the DNA channel.}
    \label{fig:fsm}
  \end{center}
\end{figure}

The DNA data storage channel is inherently error-prone due to errors in synthesis and amplification \cite{sella2021dna}, sequencing and also modifications during storage~\cite{dnaerrorchannel}. It is therefore common practice to use error correcting codes along with the data. In this work we used convolutional codes on every DNA strand individually, and not across strands. The reason we used an error correcting code only in the granularity of a DNA strand is that this is the granularity of the basecaller, i.e., the basecaller operates on a single DNA strand at a time without carrying information to or from other strands. The specific choice of convolutional codes, other than being simple to understand and implement, was motivated by the following features:
\begin{itemize}
    \item The Viterbi decoder that could be used to decode convolutional codes has the potential to recover from insertions and deletions \cite{mori1995viterbi}.
    \item Algorithms for the decoding of convolutional codes could make use of a list of probabilities per nucleotide. This soft information exists in the layer prior to last in most basecallers, where the last layer takes the argmax, i.e., the highest probable choice between A,C,T,G.
    \item Using our latest work on constraint driven encoding \cite{sella2021dna}, not only is the Viterbi decoder suitable for decoding, but the introduction of constraints on the input data has the potential to assist in reducing basecaller size.
\end{itemize}
We should state that we have yet to use these features in our work, and we expect their future utilisation to amplify our results.
A convolutional encoder can be modelled as a finite state machine, described by a list of states, actions, transitions, outputs, and a starting state. We implemented a finite-state-machine-based encoder, which processes the input stream as a list of actions. Upon receiving an action, or input bit(s), the encoder decides the next state and output based on the current state and action. The encoded stream is then obtained by concatenating the output produced by the finite state machine. The finite state machine for our 1/3 rate code is shown in Figure \ref{fig:fsm}. Similarly, we have implemented 1/2 and 2/3 rate codes, all could be found in our repository cite[removed to maintain author anonymity].
The Viterbi algorithm is commonly used as a decoder for convolutional codes and any coded sequence $\{c_0, c_1,...\}$ corresponds directly to a path through the encoder states. However, noise in the DNA channel means that when basecalling the decoder instead receives a noisy sequence $\{r_0, r_1,...\}$, which may not correspond directly to a path in the state diagram. The decoder attempts to find the maximum likelihood path with respect to the received sequence. It does so by exploring multiple possible paths and keeping track of a path metric to determine the maximum likelihood path.

\subsection{Characterisation of the DNA Channel}
To understand error rates and characteristics of Nanopore sequencing, we analyse multiple, publicly available datasets~\cite{Taiyaki/walkthrough.rstNanoporetech/taiyaki} of raw nanopore sequencing data consisting of 50k reads. The dataset contains sequences of length ranging from 1035 bases to 16909 bases. The entire data set of size $10$ GB  took us over $3.5$ hours to basecall using the Bonito basecaller. Following basecalling we sampled $200$ reads and aligned them to their reference input. We compute the rate of errors by comparing the input sequence with the output of the Bonito basecaller. We focus on three types of errors:
\begin{itemize}
    \item Insertion errors, often caused by signals from bases that are basecalled into too many bases.
    \item Deletions which are contracted signals, i.e., signals representing multiple bases are basecalled into fewer bases.
    \item Substitutions which are signals from one base that are misclassified by the basecaller.
\end{itemize}   
We compute the insertion, deletion, substitution and total error rates. The results are summarised in Table \ref{tab:taiyaki_error_summary}. We use these error characteristics to simulate errors in synthetic, encoded  data.

\begin{table}[h]
    \centering
    \caption{Average error summary for real sample data.}
    \begin{tabular}{ll}
         Mean insertion rate (\%) & 0.698 \\
         Mean deletion rate (\%) & 2.240 \\
         Mean substitution rate (\%) & 1.736 \\
         Mean total error rate (\%) & 4.674 \\
    \end{tabular}
    \label{tab:taiyaki_error_summary}
\end{table}

\begin{figure}[h]
  \begin{center}
    \includegraphics[width=.6\columnwidth]{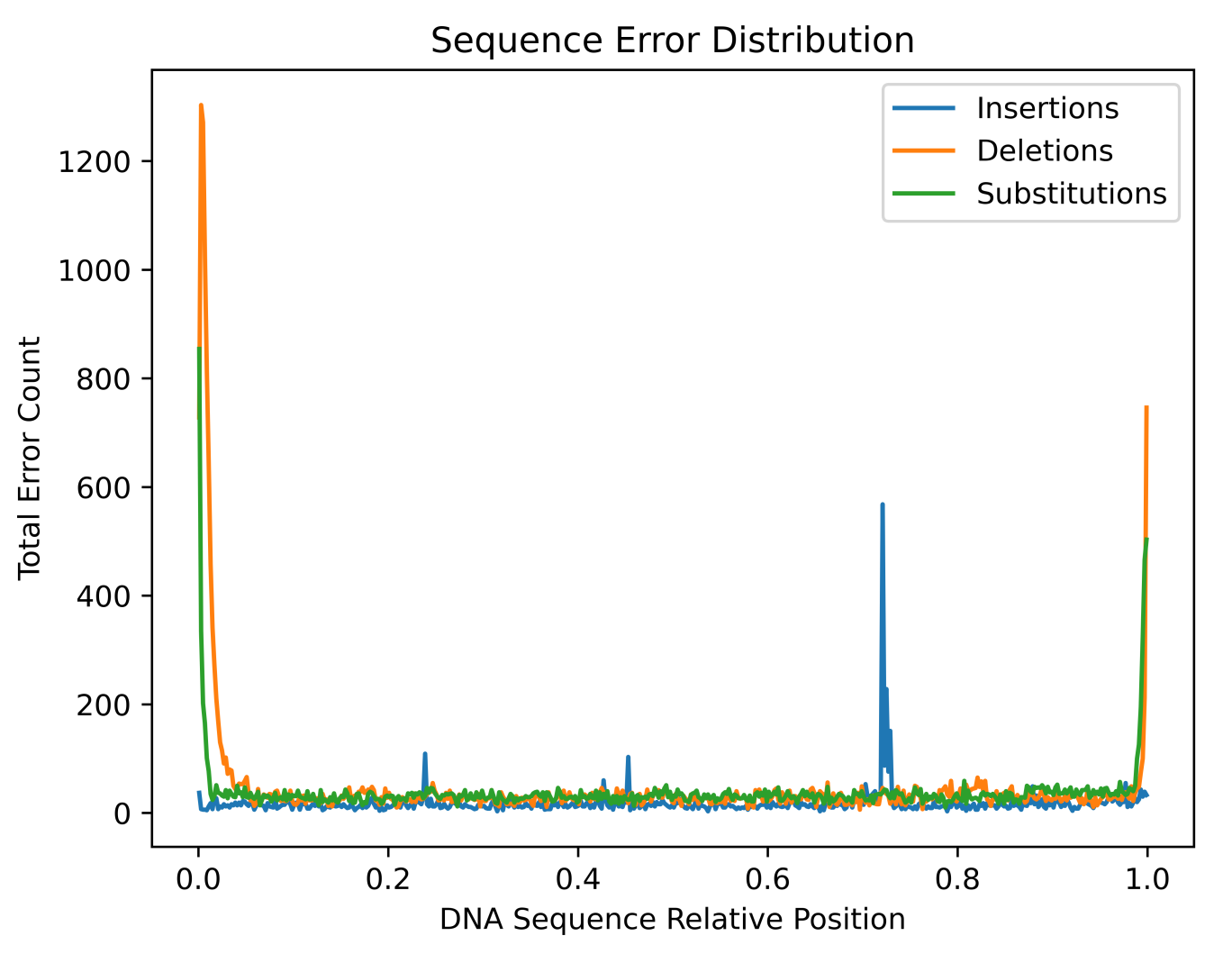}
    \caption{Distribution of errors from $200$ sampled sequences. Since the sequences sampled have variable lengths we present the position relative to the normalised sequence length.}
    \label{fig:errorModalities}
  \end{center}
\end{figure}

To further understand the distribution of errors in Nanopore sequencing, we plot the error count of each type as a function of its index in the sequence in Figure~\ref{fig:errorModalities}.

Figure~\ref{fig:errorModalities} suggests that deletions in particular occur at the beginning and end of the sequence. This is further confirmed by Table~\ref{tab:K}, where error rates of the same sequences with the first and last $K$ bases removed, for different values of $K$, are summarised.

\begin{table}[h]
    \centering
    \caption{Average error summary with first and last K bases removed.}

    \begin{tabular}{p{1cm}C{1.5cm}C{1.5cm}C{1.5cm}C{1.5cm}}
    \toprule
K &Insertion rate (\%) &Deletion rate (\%) &Substitution rate (\%) &Error rate (\%)\\
\midrule
0 &0.698 &2.240 &1.736 &4.674\\
5 &0.692 &2.002 &1.523 &4.217\\
10 &0.690 &1.940 &1.456 &4.095\\
20 &0.682 &1.813 &1.369 &3.864\\
30 &0.667 &1.657 &1.341 &3.666\\
\bottomrule
\end{tabular}
    \label{tab:K}
\end{table}

The phenomena of higher inaccuracy at the beginning and end of a sequence can sometimes be compensated for in Bioinformatics by sequencing a sliding window of a given DNA sample. In data-carrying DNA this sliding window method could be introduced by encoding the original data into multiple overlapping strands of DNA. This was introduced in \cite{Goldman2013TowardsDNA} and later coined in \cite{bornholt2016dna} as Goldman coding. Higher overlap between strands of DNA means more redundancy, and although it is not the main subject of this work, we felt it was appropriate to present our probing into the overlapping factor, i.e., K, in Nanopore sequencing.


\section{Model Compression}
\label{modelCompression}
Practically any data storage system takes into account errors and compensates for it using error correcting codes. For this reason the model compression techniques considered here could be characterised as "lossy", i.e., they are prone to reduce accuracy and thus increase errors.

Model pruning is an effective method of compression exploiting redundancy in parameters \cite{blalock2020state}. It is motivated by the possibility that neural networks may be over-parameterised, often having multiple features encoding similar information. The basic idea is that neurons in a network are ranked according to some saliency metric indicating how much they contribute to the final output. The lowest contributing neurons are then pruned away, resulting in a smaller and faster network. With a target compression ratio in mind there are several ways to proceed.

\subsection{Unstructured Pruning}

\subsubsection{One-shot Pruning}
We use the one-shot technique as described in Algorithm \ref{alg:oneshot}. x\% describes the desired compression ratio for the layers being pruned, and represents the weight-pruning threshold value. It is set to a value such that x\% of model weights lie below it.

\begin{algorithm}[h!]
\SetAlgoLined
\SetKwRepeat{Repeat}{repeat}{until}

Load pre-trained model $C$\;

    Evaluate importance of model weights using L1 norm\;
    Update $C$'s model weights mask $M$ to remove $x\%$ of least important weights:\
    
    \Indp\begin{math}
        M_{i,j} = 
            \left\{
                \begin{array}{ll}
                    0 & \mbox{if $|W_{i,j}| < \lambda$} \\
                    1 & \mbox{otherwise}
                \end{array}
            \right.
    \end{math}\;
\Indm
Merge mask $M$ into model weights $W$ to form pruned model $C'$\;
\For{i in epochs}{
    Finetune pruned model $C'$ on training data\;
}
 
 \caption{One-shot Pruning and Model Finetuning}
 \label{alg:oneshot}
\end{algorithm}

\subsubsection{Iterative Pruning}
We use the iterative pruning compression technique detailed in Algorithm~\ref{alg:iterativepruning}. As before, x\% indicates the desired compression ratio for the layers being pruned, while $\lambda$ denotes the weight-pruning threshold value. Unlike the one-shot pruning algorithm, however, the value of $\lambda$ will change with each pruning iteration, removing increasingly more parameters from the initial model. We achieved compression ratios of $(1-0.5^n)\times100\%$, where $n$ is the number of pruning iterations, for values up to n = 8, resulting in a maximum compression ratio of 99.6\%, using this technique.

\begin{algorithm}[h!]
\SetAlgoLined
\SetKwRepeat{Repeat}{repeat}{until}

Load pre-trained model $C$\;
\Repeat{$x\%$ compression reached}{
    Evaluate importance of model weights using L1 norm\;
    Update $C$'s model weights mask $M$ to remove 50\% of least important weights:\
    
    \Indp\begin{math}
        M_{i,j} = 
            \left\{
                \begin{array}{ll}
                    0 & \mbox{if $|W_{i,j}| < \lambda$} \\
                    1 & \mbox{otherwise}
                \end{array}
            \right.
    \end{math}\
    
    Train $C$ on 10\% of training data\;
}
Merge mask $M$ into model weights $W$ to form pruned model $C'$\;
\For{i in epochs}{
    Finetune pruned model $C'$ on training data\;
}
 
 \caption{Global Iterative Pruning and Model Finetuning}
 \label{alg:iterativepruning}
\end{algorithm}

An additional hyperparameter, determining the share of model parameters that are pruned in each iteration, needs to be specified for iterative pruning. The bigger this hyperparameter's value, the faster a compressed model can be created, but the poorer its accuracy is in general. Setting a small number, on the other hand, takes longer to attain a specific compression ratio, but frequently results in more accurate models. We decided to prune by 50\% each iteration, as we found empirically that this amount was not too large to result in a significant drop in accuracy after pruning, but not too small that we would need several pruning iterations to achieve our desired compression ratios.

\subsection{Structured Pruning}
Structured pruning limits pruning to patterns that result in deleting matrix columns, rows, or blocks as opposed to unstructured pruning where pruned neurons are not constrained. For structured pruning, we prune the columns of each weight matrix with the lowest L1 norm. The L1 norm of a vector is defined as $|x_1| = \sum_{i=1}^{n} |x_i|$. In our experiments we perform iterative pruning only (and not one-shot pruning).

\section{Related Work}
Nanopore sequencers generate a signal modulated by DNA molecules. They are  portable and fit for real-time sequencing \cite{BozaDeepNano-blitz:Sequencers}. Guppy, the default basecaller provided by Oxford Nanopore Technologies (ONT), offers two basecaller models: a fast basecaller and high accuracy basecaller. The fast basecaller can perform basecalling in real time, whilst the high accuracy basecaller produces higher accuracy, but cannot be used in real time. Other basecallers such as Chiron and Bonito have also been reported to be too slow for real time basecalling \cite{PeresiniNanoporeEdge}. DeepNano-blitz \cite{BozaDeepNano-blitz:Sequencers} and DeepNano-coral \cite{PeresiniNanoporeEdge} aim to address this issue. Both successfully attain real-time performance. DeepNano-blitz achieves improvement through the use of Rust and handcrafted optimisations, including cache-aware memory layouts, sigmoid/tanh approximations and the Intel MKL library for matrix multiplication.
Fast-Bonito \cite{Xu2020FastBonitoSequencing} also tackles performance issues with current basecallers through the use of knowledge distillation coupled with Neural Architecture Search (NAS) on the Bonito basecaller. 

More basecallers are being developed, however, the focus is on increasing accuracy --- as required by biological applications --- which results in deeper and more complex network architectures.


\section{Experimental Results}

\subsection{Model Analysis}

\begin{table}[!ht]
\centering
\resizebox{0.6\columnwidth}{!}{
\begin{tabular}{p{1cm}C{1.5cm}C{1.8cm}C{1.5cm}C{1.5cm}}
\toprule
\textbf{Layer} & \textbf{CPU Execution [ms]} & \textbf{CUDA Execution [ms]} & \textbf{CUDA Memory [MB]} & \textbf{Number of Parameters} \\ \midrule
Conv1 & 27.147 & 27.057 & 54.93 & 24 \\
Conv2 & 23.971 & 23.906 & 10800 & 336 \\
Conv3 & 21.688 & 83.797 & 10380 & 234240 \\
LSTM1 & 1898 & 6170 & 9480 & 4724736\\
LSTM2 & 1895 & 6153 & 9480 & 4724736 \\
LSTM3 & 1893 & 6162 & 9480 & 4724736 \\
LSTM4 & 1895 & 6151 & 9480 & 4724736  \\
LSTM5 & 1892 & 6162 & 9480 & 4724736 \\
Linear & 11.173 & 1004 & 41380 &  3937280\\ \bottomrule
\end{tabular}
}
\caption{Analysis of the different layers using Torchprof for execution times, memory and number of parameters.}
\label{tab:torchprof_analysis}
\end{table}

We begin by profiling the distribution of parameters across the model, time spent and memory usage of the Bonito model layers. 
The Bonito model essentially has three types of layers which have a substantial number of parameters: Conv1-3 (234,240 parameters), LSTM1-5 (4,724,736 parameter each) and Linear (3,937,280 parameters).

An analysis of the distribution of parameters across the layers of the network shows that the model's high parameter count can be attributed primarily to the weights of the LSTM layers and linear layer. The weights of the convolutional layers have little effect on the parameter count, and the biases of each layer account for a negligible proportion of the parameters. Our focus thus lies on compressing the weights in the LSTM and linear layers. 

Next we used Torchprof~\cite{torchprof} to perform layer-by-layer profiling of the Bonito model, revealing statistics on memory utilisation and latency of each individual layer in the network. The results, presented in Table~\ref{tab:torchprof_analysis}, corroborate the same conclusion as our examination of the parameter distribution among network layers: during inference, the layers with the biggest memory and latency overhead are the LSTM and linear layers (see Table~\ref{tab:torchprof_analysis}) near the network's end.
According to the CPU Execution and CUDA Execution columns, the LSTM and linear layers account for 99.2\% of the latency overhead in prunable layers on CPU and 99.6\% on GPU. The results of CUDA Memory also shows that memory usage in these layers is high compared to convolutional layers. As a consequence, we compress the weights of these layers.

\subsection{Methodology}
We use the default pre-trained Bonito model as a starting point. We trained Bonito on 66k reads from DNA of E. coli, H. sapiens, and S. cerevisiae (yeast). For training, the readings are chunked, and the model is trained with a CTC-based loss. For the assessment of the model, a second validation set of 1000 readings is employed. The original model was trained using the same train-test split.

As discussed in \textbf{Model Compression}, we used different pruning approaches. We pruned neurons in an unstructured way using global unstructured pruning. We initially tested one-shot pruning, which tries to achieve the required compression ratio in one single pruning step. This has the advantage of not requiring extra hyperparameters or a pruning schedule; rather, the model is trimmed once, then retrained. We then test iterative pruning  in which model parameters are gradually trimmed away in pruning stages interspersed with retraining stages. Doing so has the potential to lead to a better model performance and to better compression.
Finally, we tested structured pruning, which entails removing weights according to a predetermined sparsity pattern.

Based on the findings of our profiling, we prune the weights in the LSTM and linear layers of the model --- the layers which have the most parameters and latency overhead --- to a predetermined compression ratio. We use magnitude-based pruning, in which the L1 norm, or its absolute size, is used as the saliency metric for ranking network weights. 
When retraining the model post pruning, hyperparameters such as learning rate and training epochs are set based on the loss observed during preliminary model training experiments. To obtain a constantly diminishing loss while the model learns, we train for five epochs with a learning rate of 0.0005. The results of our experiments are in Table~\ref{table:globaliterativepruning}. For the model size we report the size on disk which depends somewhat on the implementation.



\begin{table*}[!ht]
\centering
\caption{Mean accuracy due to compression for Unstructured (Iterative), Unstructured (One-shot) and Structured pruning.}
\label{table:globaliterativepruning}
\resizebox{\textwidth}{!}{
\begin{tabular}{R{2.9cm} R{2cm} R{1.75cm} | R{2.5cm} R{2.5cm} R{2.5cm}} \hline
\multicolumn{1}{L{2.9cm}}{\textbf{Compression Ratio (\%)}} & \multicolumn{1}{L{2cm}}{\textbf{Parameters Removed (\%)}} & \multicolumn{1}{L{1.75cm}|}{\textbf{Stored Model Size (MB)}} & \multicolumn{3}{C{8.5cm}}{\textbf{Mean Accuracy (\%)}} \\ \cline{4-6}
& & & \multicolumn{1}{L{2.5cm}|}{ } & \multicolumn{1}{L{2.5cm}|}{ } & \multicolumn{1}{L{2.25cm}}{ }\\
& & & \multicolumn{1}{L{2.5cm}|}{\textbf{Unstructured, iterative}} & \multicolumn{1}{L{2.5cm}|}{\textbf{Unstructured, one shot}} & \multicolumn{1}{L{2.25cm}}{\textbf{Structured}} \\ \hline   
0 & 0 & 107 & 97.371 & 97.205 & 97.371\\
50 & 49.541 & 107 & 97.245 & 97.263 & 93.525\\
75 & 74.311 & 107 & 97.194 & 97.201 & 88.564\\
87.5 & 86.697 & 73 & 96.871 & 96.842 & 76.224\\
93.75 & 92.889 & 40 & 96.330 & 95.434 & 64.131\\
96.875 & 95.985 & 23 & 95.257& 92.257 & 48.041\\
98.4375 & 97.534 & 15 & 93.209& 89.314 & 8.727 \\
99.21875 & 98.308 & 11 & 90.668 & 83.228 & 0.194\\
99.609375 & 98.695 & 8.6 & 87.016 & 71.813 & 0
\end{tabular}
}
\end{table*}


\subsection{Accuracy vs Compression Comparison}
\begin{figure}[!ht]
  \begin{center}
    \includegraphics[width=.6\columnwidth]{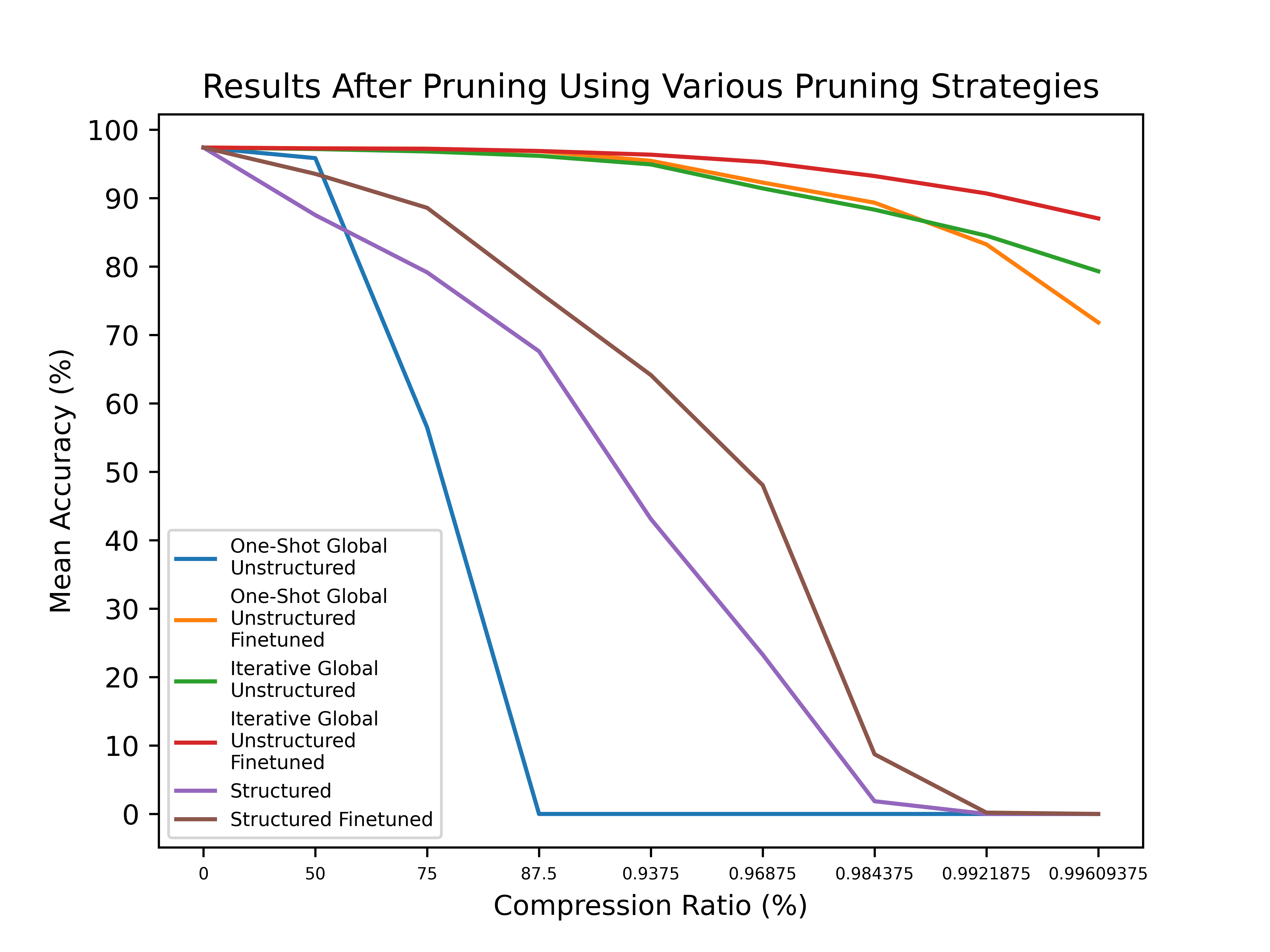}
    \caption{Performance of our pruned models after pruning using all the strategies we experimented with. Our proposed strategy, iterative global unstructured pruning, is illustrated in red.}
    \label{fig:comparisonaccuracy}
  \end{center}
\end{figure}

We compare the techniques we have tested with each other in  Figure~\ref{fig:comparisonaccuracy} based on the accuracy and the model compression.

Table~\ref{table:globaliterativepruning} and Figure~\ref{fig:comparisonaccuracy} indicate that when using structured pruning, we are unable to prune to the same amount as when using unstructured pruning, and that models reduced to the same degree of sparsity when pruned in a structured way have considerably worse accuracy than when pruned unstructured. This is likely due to our structured pruning technique being applied separately to each layer, i.e., we prune all layers evenly rather than globally ranking the weights across all layers to be pruned as we did in case of unstructured pruning. Because the layers in our model have various sizes, implementing global structured pruning is difficult. For example, ordering the columns of weight matrices according to their L1 norm makes no sense when some columns have more elements than others. Future research might be based on further investigation of how to accomplish this successfully or exploring different trimming techniques.

Our results generally suggest that iterative pruning outperforms one-shot pruning for our model, with up to 15.2\% higher accuracy at a compression ratio of 99.6\%. This is most likely because pruning too much at once means the model can no longer recover, even after retraining. Iterative pruning, on the other hand, allows the model to be trained while being pruned, allowing the remaining weights in the model to adapt to compensate for those that were lost during pruning. 

\begin{figure}[!ht]
  \begin{center}
    \includegraphics[width=.8\columnwidth]{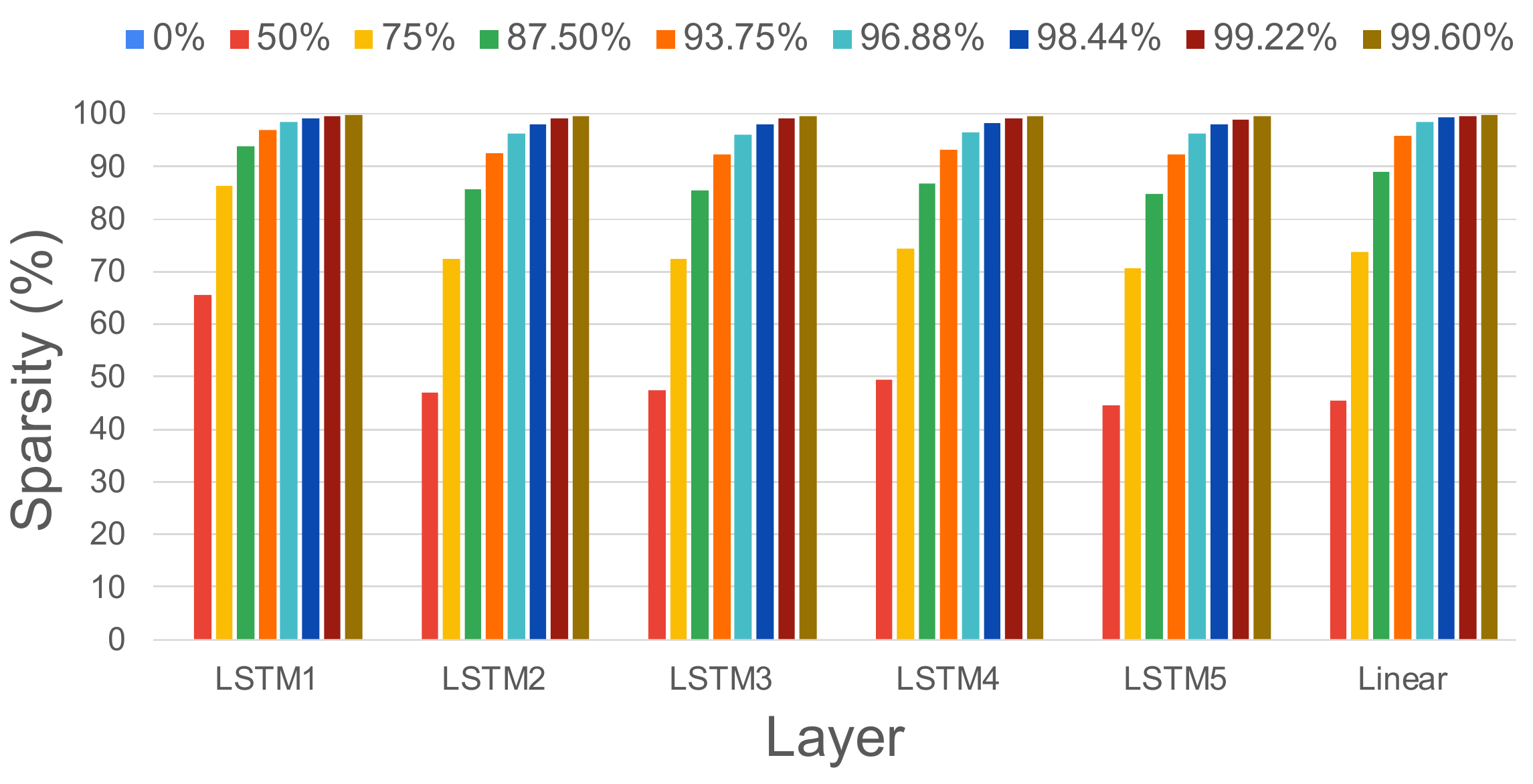}
    \caption{Sparsity distribution in model layers after iterative global unstructured pruning.}
    \label{fig:sparsity}
  \end{center}
\end{figure}

Model weights may not necessarily be eliminated in equal proportions from each layer when using global pruning, e.g., deeper network levels may be pruned more extensively, allowing for more connections from deeper layers to be eliminated. Our analysis summarised in Figure~\ref{fig:sparsity} shows, however, that the initial LSTM layer is obviously pruned more than other layers with comparable sparsity at lower compression ratios. On the other hand, at extremely high compression ratios, they all have similar sparsity.

\subsection{Notes on Model Speedup}
Pruning models in PyTorch, the framework used for Bonito, improves efficiency by using sparse tensor representations. There is, however, still little support in deep learning frameworks such as PyTorch or in general purpose hardware for sparse tensor operations. While support is increasingly being added, and PyTorch is striving to improve the complexity of sparse tensor operations to match those of other libraries such as numpy~\cite{numpy}, critical aspects such as batch data management and backpropagation are currently unsupported. Taking advantage of model speedup will therefore require greater support for sparse tensors and sparse networks, or implementation on special purpose hardware.

\begin{figure}[ht!]
  \begin{center}
    \includegraphics[width=.6\columnwidth]{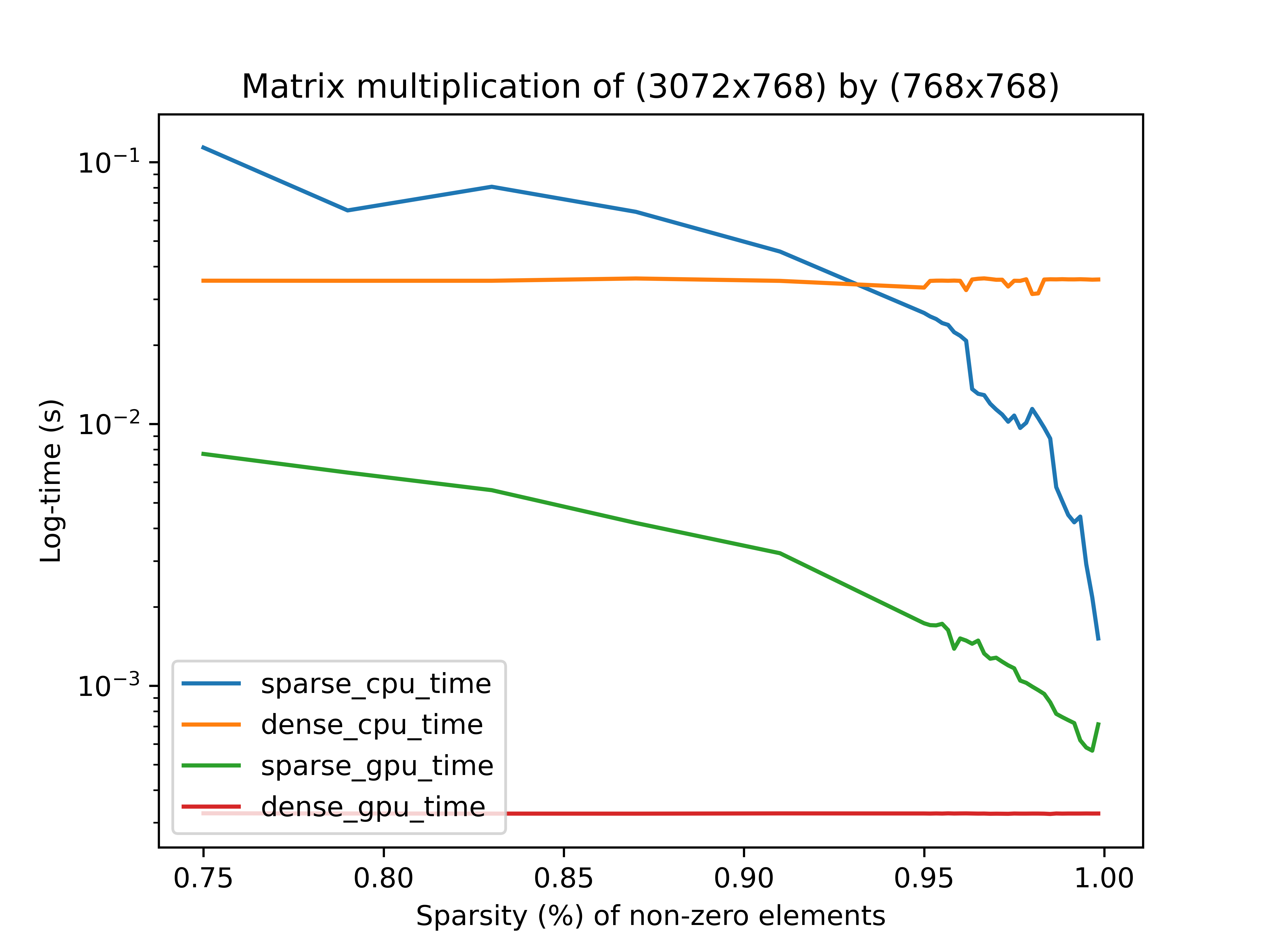}
    \caption{Illustration of the speed of sparse-dense matrix multiplication for tensors of varying sparsity in PyTorch.}
    \label{fig:comparisonspeed}
  \end{center}
\end{figure}

PyTorch presently uses the sparse coordinate list storage format, which consists of a list of non-zero indices and a list of values at those indices. Other sparse tensor layout types include compressed sparse row and compressed sparse column, which can perform better for arithmetic and matrix multiplications. To offer some insight into the current state of sparse tensors in PyTorch, we investigated the relationship between the sparsity of a (3072x768) tensor (the size of weights in a Bonito LSTM layer) and the time required for matrix multiplication. Figure~\ref{fig:comparisonspeed} summarises the results.

We found that when the sparsity is larger than 93\%, the overhead of using sparse tensors is less than the performance savings from using sparse tensors, and so a considerable speedup is realised by using a sparse representation rather than the usual dense representation. We obtain a 10$\times$ speedup on CPU at a sparsity of 99.6\% (the same as our most pruned model).

On GPU, the size of our LSTM weight matrix is insufficient to benefit from any computational speedup using sparse tensors. If the weight matrix were of a larger dimension, there would be a point at which matrix multiplication performs better on a sparse tensor. However, as mentioned previously, improving the efficiency of sparse tensor operations in PyTorch is work in progress, and this graph may see significant improvements in future releases.

\section{Compensating for Accuracy Loss \label{sec:compensatingForAccuracy}}
Our work on pruning shows that we are able to significantly compress the Bonito basecaller model with some tradeoff to model accuracy. In fact, pruning away 98.7\% of the model results in a 10.4\% drop in accuracy, implying that 98.7\% of the model is used to achieve the last 10.4\% of accuracy. At the same time, error correction codes have the potential to pick up some of the errors resulting from model compression. We first analyse how many errors can be fixed by using convolutional codes and then study how we can use error correction codes in the context of model compression.

\subsection{Potential of Error Correction}
To understand the potential of convolutional error correction codes for the DNA channel in general, we investigate it with three different code rates 1/3, 1/2 and 2/3 on sample data.

We use a snippet of \textit{A Tale of Two Cities} by Charles Dickens~\cite{Dickens1994ACities} as sample data of 8.2KB. We convert the text into binary chunks before converting each chunk into a DNA sequence. We run an error simulator \cite{ErrorSim} on the sequences, simulating substitution errors only. To read the data, we convert the sequences to binary and then to text. 

To measure the error rate, we compare the DNA version of the data after decoding with the DNA version of the original data, performed Smith-Waterman alignment~\cite{SMITH1981195}
and measured the error rate. We measure errors at the DNA-level rather than at the binary-level, as our errors occur at the DNA-level and measuring errors at the DNA-level thus avoids a misleading error rate, since a single DNA substitution can correspond to either 1 bit flip, or 2 bit flips in 33\% of cases.
We test two DNA sequence lengths, 150 and 300 nucleotides (nt), as these are available commercially. 

The results of our experiments summarised in Table~\ref{tab:code_rate_vs_error_rate} show that both the 1/2 and 1/3 codes are able to significantly reduce the error rate of the data read back from the baseline, improving the error rate by up to 93.4\% for a code rate of 1/3 and DNA sequence length of 300nt. Using the rate 2/3 code, however, increased the error rate. This is attributed to decoder failure, i.e.: there are too many errors for the decoder to compensate for, often resulting in an increased error rate.

\begin{table}[!ht]
\centering
\resizebox{0.6\columnwidth}{!}{
\begin{tabular}{p{1.5cm}C{2.25cm}C{2.9cm}C{2.25cm}}
\toprule
\textbf{Code Rate} & \textbf{Sequence Length (bases)} & \textbf{Number of DNA Sequences} & \textbf{Error Rate (\%)} \\ \midrule
None & 150 & 224 & 2.696 \\
None & 300 & 112 & 1.507 \\
2/3 & 150 & 335 & 5.228 \\
2/3 & 300 & 168 & 2.817 \\
1/2 & 150 & 447 & 1.082 \\
1/2 & 300 & 224 & 0.472 \\
1/3 & 150 & 670 & 0.403 \\
1/3 & 300 & 335 & 0.099 \\ \bottomrule
\end{tabular}
}
\caption{Error correction potential of different code rates for different sequence lengths.}
\label{tab:code_rate_vs_error_rate}
\end{table}

On DNA sequences of length 300nt, Using a 1/2 code reduces the error rate from 1.507\% to 0.4723\%, whilst using the 1/3 code reduces the error rate to 0.0988\%, corresponding to an improvement of 68.7\% and 93.4\% respectively.

\begin{figure}[ht]
  \begin{center}
    \includegraphics[width=.6\columnwidth]{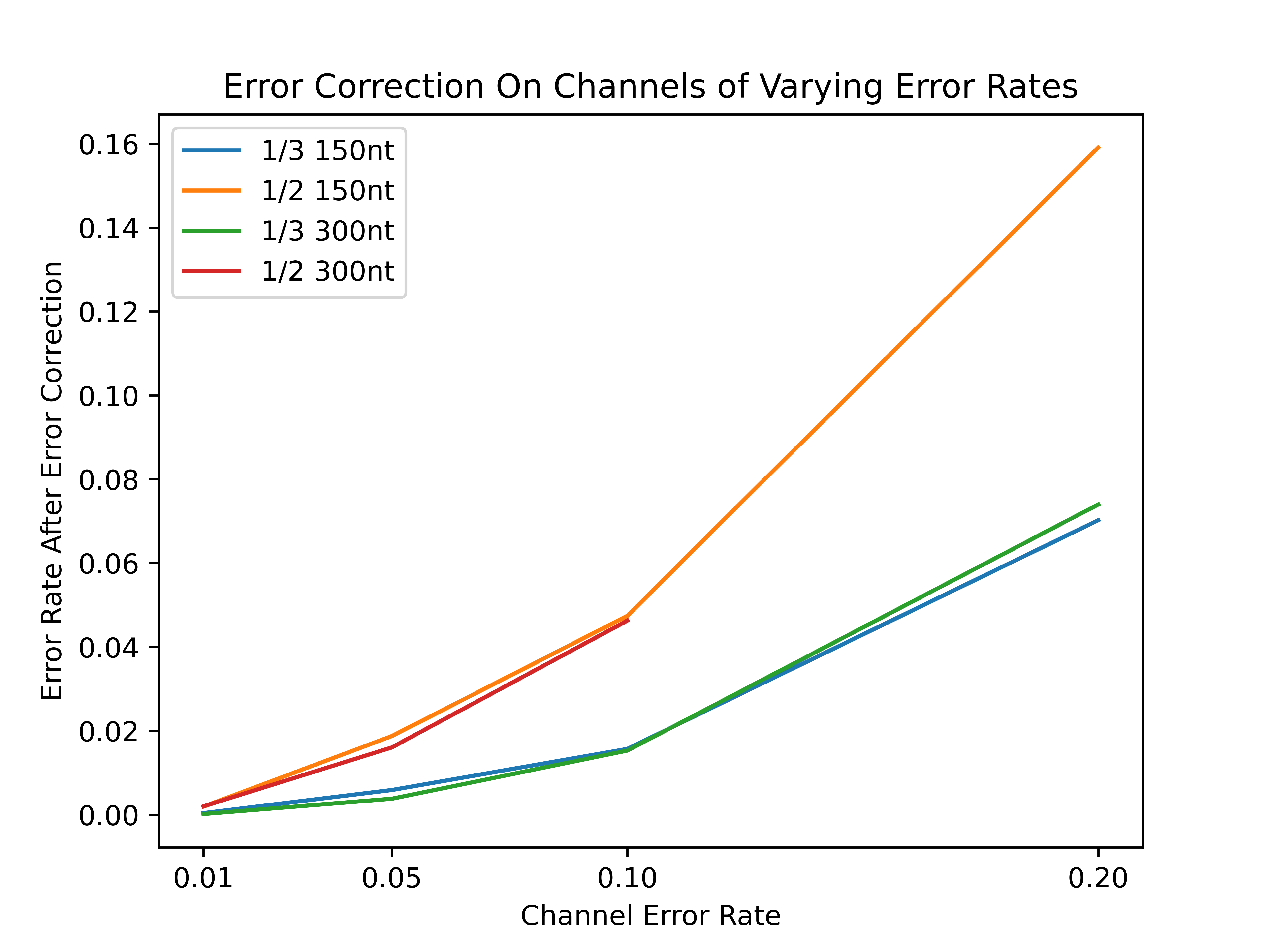}
    \caption{Error correction capability for 1/2 and 1/3 code rates.}
    \label{fig:error_correction_capability}
  \end{center}
\end{figure}

To further investigate the error correcting capability of both codes, we configure our error simulator to simulate error rates of 1\%, 5\%, 10\% and 20\% on the data, and plot the error rate of the data read back in Figure~\ref{fig:error_correction_capability}). 

For the lowest error rate of 1\%, using the 1/3 code on DNA sequences of both length 150 and 300 bases resulted in the majority of DNA sequences read back error free, with error rates of 0.0389\% and 0.0180\% respectively. This represents an improvement of 96.1\% and 98.2\% respectively. A similar proportion of errors are corrected for an error rate of 5\%. At higher error rates, a smaller proportion of errors are corrected, but a considerable reduction in errors is still seen at error rates of 10\% and 20\%, where the error rate falls to 1.530\% and 7.022\% after error correction with a 1/3 code. 

\subsection{Compensation Potential}
Our investigation into a convolutional error correction code shows it is successful at reducing error rate by up to 13.0\% on sample data, leading to the question: what if instead of using the extra 98.7\% of the model to achieve high accuracy, we can use an error correction code to pick up the errors generated by model compression, and maintain the performance gains from using the compressed model?
To answer this we encoded the same data using our 1/2 and 1/3 rate encoding and put it through our error simulator \cite{ErrorSim} using our iteratively-unstructured-pruned models rather than the original pre-trained Bonito model, replicating substitution errors. We calculated the error rate after decoding and compared it to the error rate without any error correcting code. Table~\ref{tab:model_compression_w_ecc} summarises our findings.

\begin{table}[ht!]
\centering
\setlength{\tabcolsep}{5pt}
\resizebox{0.7\columnwidth}{!}{
\begin{tabular}{p{2.5cm}C{2.3cm}ccc}
\toprule
\textbf{Compression ratio (\%)} & \textbf{Parameters Removed (\%)} & \multicolumn{3}{C{4.5cm}}{\textbf{Error Rate (\%)}} \\
\cmidrule{3-5}
& & \shortstack[l]{\textbf{No} \\ \textbf{Code}} & \shortstack[l]{\textbf{1/2 Rate} \\ \textbf{Code}} & \shortstack[l]{\textbf{1/3 Rate} \\ \textbf{Code}} \\
\midrule
0 & 0 & 1.507 & 0.490 & 0.099 \\
50 & 49.541 & 1.715 & 0.496 & 0.210 \\
75 & 74.311 & 1.408 & 0.463 & 0.123 \\
87.5 & 86.697 & 1.481 & 0.463 & 0.150 \\
93.75 & 92.889 & 1.811 & 0.443 & 0.186 \\
96.875 & 95.985 & 2.153 & 0.409 & 0.165 \\
98.4375 & 97.534 & 3.784 & 1.235 & 0.467 \\
99.21875 & 98.308 & 5.985 & 2.505 & 0.773 \\
99.609375 & 98.695 & 11.565 & 2.948 & 2.225 \\ \bottomrule
\end{tabular}
}
\caption{Results of applying error correction in conjunction with model compression.}
\label{tab:model_compression_w_ecc}
\end{table}

Our results show that when using our second and fourth-most compressed models, we can reduce the error rate from 5.985\% to 0.773\% and from 2.153\% to 0.165\% by applying the 1/3 code. When compared to the original Bonito model and error rate, this is a 98.308\% and 95.985\% reduction in model size, evaluated in terms of parameter count, and a 48.739\% and 89.071\% improvement in error rate. Furthermore, up until our third-most compressed model, we achieved error rates of 0.5\% using both 1/2 and 1/3 coding rates. This is a significant result, as we can conclude that we are able to drastically reduce basecaller model size and still gain a lower error rate through the use of error correcting codes.

\section{Conclusions and future work}
DNA data storage is a likely contender to be the next archival storage medium but before this happens a number of challenges need to be addressed. 
One such challenge is the accuracy and computational burden of the machine learning models used for basecalling, i.e., during the process of reading the information back.
The heavy lifting in both write and read side are still done by methods designed for life sciences, which do not take into account our control over how data is represented in DNA, and anything else we can do to improve read accuracy.

In this paper we investigate "lossy" model compression of the basecalling network, resulting in smaller size and with the potential of accelerating basecalling. At the same time we show a path to compensate for said lost accuracy, ensuring high-fidelity data retrieval from DNA storage through the addition of error correcting codes.

Most importantly, we show in Table~\ref{tab:model_compression_w_ecc} that the combination of a compressed model with error correcting codes can deliver higher accuracy retrieval of data than an uncompressed model without any error correction.

Based on the insights on compensating for model compression described in Section~\ref{sec:compensatingForAccuracy}, as well as the error distribution prior to model compression shown in Figure~\ref{fig:errorModalities}, our ongoing work investigates further fusion between embedded alignment markers in the data and basecallers, processing of data-carrying DNA in the signal domain (also known as "squiggle") and applying further information theoretic tools with existing sequencing methods. In particular we intend to leverage the soft information from the Bonito basecaller to obtain better error correction.




\bibliographystyle{abbrv}
\bibliography{references}

\end{document}